# Photonic Characterization of Oxygen and Air-Annealed $Zn_3N_2$ Thin Films

**Ting Wen and Ahalapitiya H Jayatissa***

Mechanical, Industrial and Manufacturing Engineering Department, The University of Toledo, Toledo, OH 43606. *Corresponding Author: E-mail: ahalapitiya.jayatissa@utoledo.edu*

**Abstract:** Zinc nitride films were synthesized by reactive radio frequency (rf) magnetron sputtering of a zinc target using an $Ar+N_2$ mixture. The as-deposited films were annealed in the air and $O_2$ at 300 °C for 1 hr. The XRD measurements indicated that the films had a polycrystalline structure with a preferred (400) $Zn_3N_2$ orientation. The annealing process enhanced the crystallinity. After annealing, the AFM and SEM morphology revealed no significant change in the surface morphology and surface roughness. The direct bandgap of $Zn_3N_2$ was estimated to be in the range of 1.15 -1.35 eV where annealing resulted in a reduction of the bandgap. The films were confirmed to be p-type conduction and the resistivity was slightly increased by annealing. The photoconductivity measurements indicated that the as-deposited films did not have any photoresponse, whereas the annealed films exhibited photoconductivity.

## 1. Introduction

Zinc nitride ($Zn_3N_2$) is a relatively least studied $A^{II}B^{V}$ type compound semiconductor. It is dark blue in colour and cubic in structure with lattice constant being $a = 9.78$ Å [1]. Numerous deposition methods have been employed to synthesize $Zn_3N_2$ films [1-10]. Widely used method is the rf-magnetron sputtering deposition, which is a very useful application in terms of large-area coating of uniform thin semiconductor layers. The reactive sputtering of zinc target in a $N_2$ and Ar gas mixture has been employed. The $Zn_3N_2$ film synthesized by rf-magnetron sputtering has a direct band gap of $1.23 \pm 0.02$ eV [5], which has strong absorption in IR region. The $Zn_3N_2$ with nanowire microstructure was reported to exhibit ultraviolet and blue emission [8]. This interesting phenomenon indicated that zinc nitride may be suitable as an optoelectronic material for infrared (IR) (750 - 1600 nm) sensors, smart window, and energy conversion devices. However, being a new material, its optical properties are still controversial, and its behaviour of photoconductivity has not yet been investigated.

According to the reviewed works, optical and electronic properties of $Zn_3N_2$ films varied over a wide range, depending upon the deposition process and nitrogen amount in the films [2-22]. The zinc and nitrogen concentration in $Zn_3N_2$ films can be controlled by the substrate temperature, $N_2$ flow concentration in the sputtering mixture and deposition time. However, the film composition could be easily changed just after deposition, because $Zn_3N_2$ is unstable and can absorb oxygen in the humid air. Heat treatment improves the electronic properties while producing more stable $Zn_3N_2$ phase. Thus, thermal annealing is used to improve the crystal quality and to remove structural defects in $Zn_3N_2$.

In this paper, the $Zn_3N_2$ films were deposited by a radio frequency (rf) magnetron sputtering method. Thermal annealing was conducted in the air and $O_2$ to stabilize the film and modify the band gap. The films were investigated for structure, optical, and electrical properties. The photoconductivity of the films for IR light was measured. The properties of the as-deposited and annealed $Zn_3N_2$ films were compared to the corresponding ones deposited for 30 min [22] to study the effect of deposition time.

## 2. Experimental

Zinc nitride films were deposited by rf magnetron sputtering system using a 6-inch Zn target (1/4-inch-thick, purity 99.995%) and non-alkaline glass substrates (0.75 mm thick). The substrates were rinsed with acetone and isopropyl alcohol before introducing to the sputtering system. The base-pressure was in the range of $2 \times 10^{-6}$ - $5 \times 10^{-6}$ Torr, and the deposition pressure was around 5 $\times 10^{-3}$ Torr. The system was purged with Ar for 30 min prior tom deposition to make the chamber oxygen-free. The target was pre-sputtered for 5 min to remove oxide layers produced in the target by exposing it to the air. Then the substrate was repositioned, and the sputtering was carried out using a mixture of nitrogen and argon (99.999%, $N_2$/Ar: 1/1). The rf power density was 2.14 W/cm$^2$ and the films were deposited for 1 hr. without heating the substrate. The as-deposited films were annealed in the air and $O_2$ respectively at 300 °C for 1 hr. using a tube furnace.

The crystallinity of the films was examined with the X-ray diffraction (XRD, Cu, K radiation, PANalytical X'Pert Pro MPD). The film surfaces were studied with atomic microscope (AFM) and scanning electron microscope (SEM). The ultra-violet (UV) and visible double beam spectrometer (UV-1650 PC Shimadzu) was used for measuring the optical transmittance and reflectance.

A thin layer of Au (~75 nm) was coated on the films to measure electrical properties and infrared sensitivity using thermal vacuum evaporation method through a shadow mask. The width and spacing of the Au electrodes were 6 mm and 4 mm, respectively. Hence, the light irradiated area was 0.24 cm$^2$. The illumination of IR light of wavelength 850 nm and intensity around 2.16 µW/mm$^2$ was provided by a light emitting diode (LED). The photoconductor was installed in a closed glass chamber (200 cm$^3$) with Au connecting electrodes and tested in vacuum ($10^{-3}$ Torr). The resistance of zinc nitride layer between Au electrodes was measured using a high mega ohm multimeter (Keithly Model: 22-816). The data were collected using a commercial software package (LabView) available from the National Instruments Company. The electric property was investigated by Van der Pauw measurement.

## 3. Results and Discussion

### *3.1 Structural property*

The XRD patterns shown in Fig. 1 characterized the crystalline structure of the as-deposited and annealed $Zn_3N_2$ films. The (222), (321), (400), (440) and (600) diffraction peaks indicated that the as-deposited and annealed $Zn_3N_2$ films had a cubic anti-bixbyite $Zn_3N_2$ structure according to the report in JCPDS file [23]. One ZnO peak with low intensity was also detected at $2\theta = 36.25°$ corresponding to (101) plane in the as-deposited film, which suggested that the as-deposited film was oxidized during storage in the ambient air prior to the XRD characterization. Comparing to the $Zn_3N_2$ films deposited for 30 min [22], the films in this study had a preferred orientation at (400) plane rather than (321) plane. It indicated that deposition time heavily affected the structural property of $Zn_3N_2$ films. Relative intensity of (400) peak was defined as,

$$i_{(400)} = \frac{I_{(400)}}{I_{(400)}+I_{(222)}+I_{(321)}+I_{(101)}+I_{(440)}+I_{(600)}} \tag{1}$$

The calculated value of relative intensity $i_{(400)}$ was shown in Fig. 2, which was much higher than the $i_{(321)}$ of the $Zn_3N_2$ films deposited for 30 min [22]. Furthermore, $i_{(400)}$ was dramatically increased and the ZnO phase was removed after annealing in the air and O₂. It suggested that more single crystalline structure was achieved for the annealed films with longer deposition time.

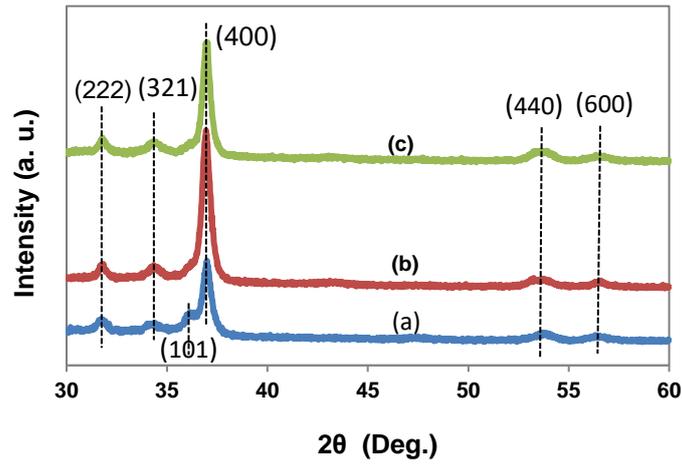

**Fig. 1:** X-ray diffraction patterns of the (a) as-deposited, (b) air-annealed, and (c) O$_2$-annealed Zn$_3$N$_2$ films surfaces.

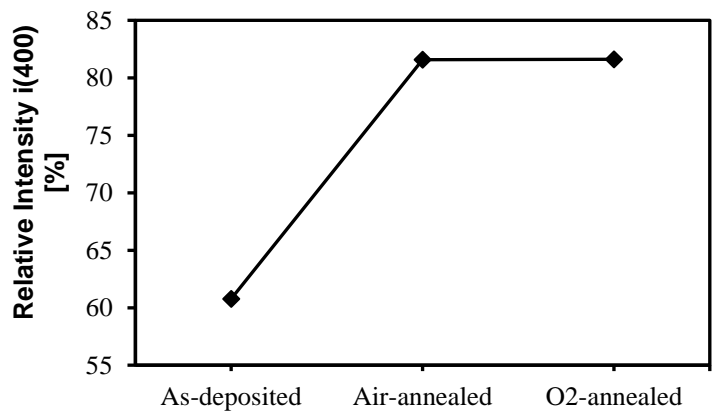

**Fig. 2:** Relative intensity of i$_{(400)}$ of the as-deposited and annealed Zn$_3$N$_2$ films

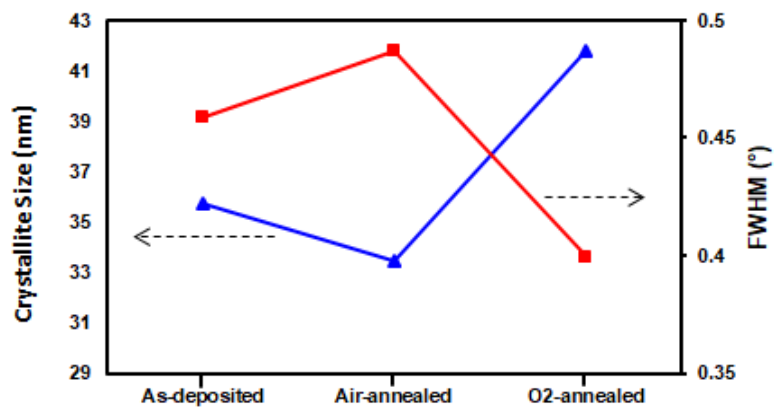

**Fig. 3:** Grain size and FWHM of the (400) phase in the as-deposited and annealed Zn$_3$N$_2$ films

The mean crystallite size of the predominant phase (400) was estimated using the Scherrer formula [24, 25].

$$D = 0.9\lambda/[(B) * \cos\theta_B]. \tag{2}$$

Where, λ, $\theta_B$ and B are the X-ray wavelength (1.5418 Å), Bragg diffraction angle and the instrumental broadening, respectively. Using $B = 0.05°$ and FWHM at (400) reflections, the grain sizes were estimated to be ≈ 35.77, 33.48 and 41.80 nm, respectively, illustrated in Fig. 3. The changes of the calculated crystalline size are also visible in the surface roughness obtained from AFM measurements (Table 2).

*3.2 Surface Property*

Figure 4 and 5 showed the surface microstructure of the as-deposited and annealed $Zn_3N_2$ films in 2 μm scale using atomic microscope (AFM) and in 5 μm scale using scanning electron microscope (SEM), respectively. The corresponding film roughness was given in Table 1. It was observed that the particle size and the surface roughness were increased comparing to that of the $Zn_3N_2$ films deposited for 30 min [22]. The grain illustrated in Fig. 4 was in consistent with the calculated crystalline size in Fig. 3. The change of surface roughness was reasonable since larger grain size resulted in rougher surface. All in all, the film roughness and particle size were not significantly changed by annealing.

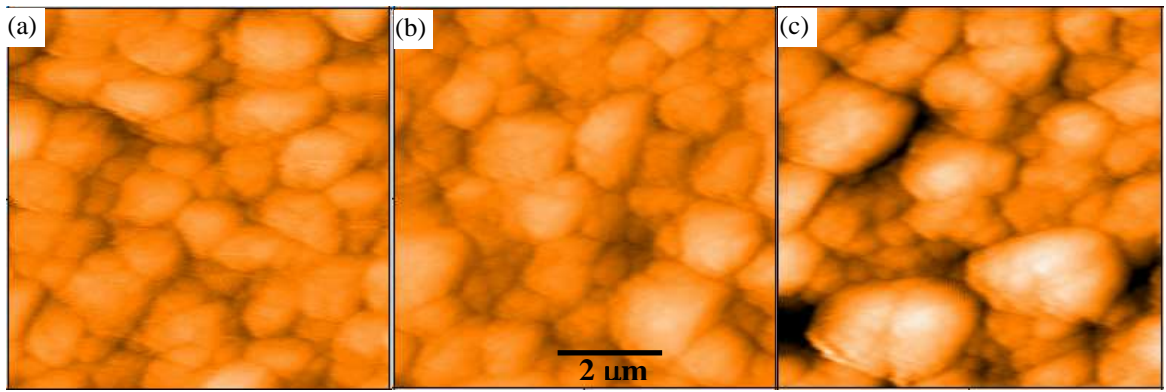

**Fig. 4:** AFM morphology of the (a) as-deposited, (b) air-annealed, and (c) $O_2$-annealed $Zn_3N_2$ films surfaces.

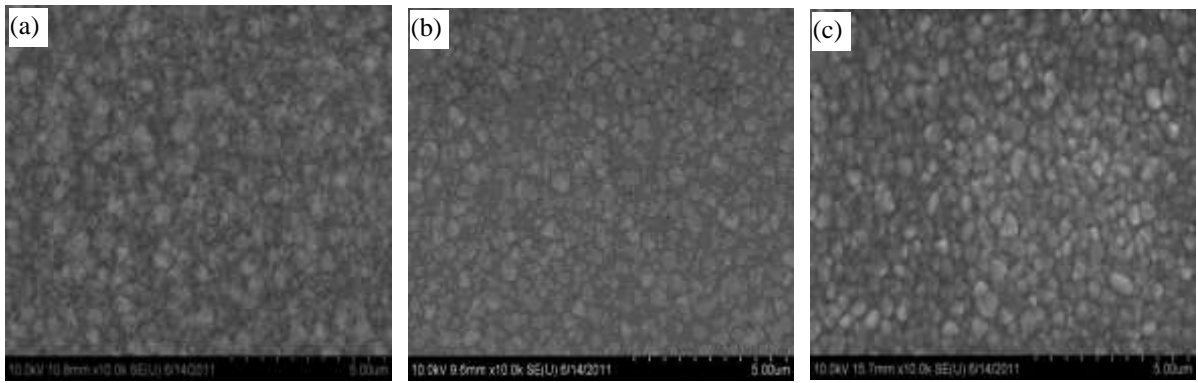

**Fig. 5:** SEM morphology of the (a) as-deposited, (b) air-annealed, and (c) $O_2$-annealed $Zn_3N_2$ films surfaces scale 5 microns).

**Table 1:** Average surface roughness of the as-deposited and annealed $Zn_3N_2$ films.

| Sample | As-deposited | Air-annealed | $O_2$-annealed |
|---|---|---|---|
| Surface Roughness (nm) | 17.0 | 15.4 | 30.2 |

## 3.3 Optical property

Figure 6 and 7 plotted the optical transmittance and reflectance of the $Zn_3N_2$ films in the photon wavelength range of 200-1800 nm. Both the transmittance and reflectance were decreased comparing to those of the $Zn_3N_2$ films deposited for 30 min [22]. It suggested that the $Zn_3N_2$ films could absorb more light as increasing the deposition time. The transmittance was decreased about 10% after annealing in the air and $O_2$. The absorption edge was shifted from 800 nm to 980 nm after annealing.

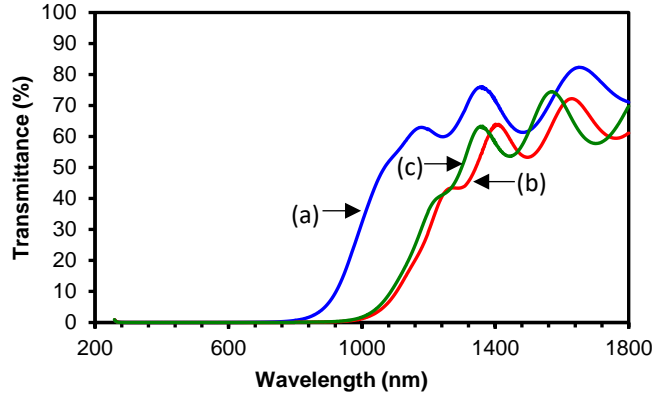

**Fig. 6:** Optical transmittance spectra of the (a) as-deposited, (b) air-annealed, and (c) $O_2$-annealed $Zn_3N_2$ films surfaces.

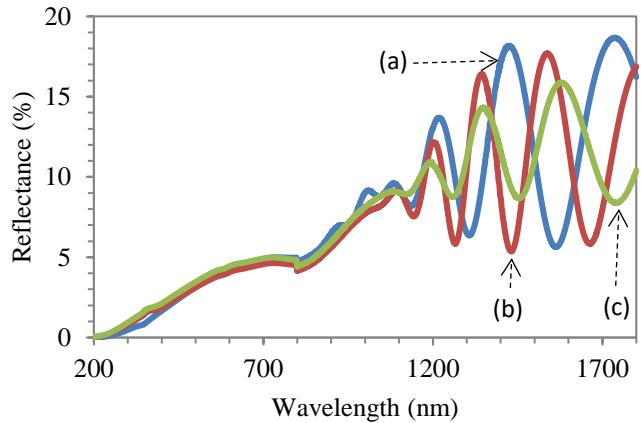

**Fig. 7:** Optical reflectance spectra of the (a) as-deposited, (b) air-annealed, and (c) $O_2$-annealed $Zn_3N_2$ films surfaces.

The thickness of $Zn_3N_2$ film is calculated with the transmittance spectrum using the formula [26,27]:

$$x = \frac{1}{2n(\frac{1}{\lambda_1}-\frac{1}{\lambda_2})} .  \qquad (3)$$

Where, x is the thickness of the $Zn_3N_2$ film, n is the refractive index, $\lambda_1$ and $\lambda_2$ are the wavelength of adjacent first two peaks from low energy side in the transmittance spectrum. By fitting of calculated spectra with the measured optical transmittance spectra. we have estimated the film thickness to be x = 2576, 2640, and 2661nm for the as-deposited, air-annealed, $O_2$-annealed $Zn_3N_2$ films, respectively. The thicknesses of the above films measured with AFM were 2560 nm, 2600 nm, and 2643 nm, respectively. The increased thickness can be interpreted as due to the improved surface crystalline structure in the (400) orientation and the interstitial oxygen ions introduced from the air and $O_2$.

The absorption coefficient, $\alpha$, was calculated using the following equation [28],

$$T = [(1 - R)^2 \exp(-\alpha d)] / [1 - R^2 \exp(-2\alpha d)]. \tag{4}$$

Where, T, R, $\alpha$, and d represents the transmittance, reflectance, absorption coefficient the film thickness respectively. Fig. 8 shows the plot of $(\alpha h\nu)^2$ versus photon energy $(h\nu)$. It is typical in direct semiconductors that the absorption coefficient reaches a value greater than $10^4$ cm$^{-1}$ at photon energies above the absorption edge. Table 2 shows the corresponding band gap change with annealing. The calculated band gaps are close to that of the corresponding $Zn_3N_2$ films deposited for 30 min [22]. The energy gap was estimated from the Tauc's plot (Fig.8) and the optical gap was estimated to the photon energy corresponding to $\alpha = 10^4$ cm$^{-1}$. Optical gap of $\alpha = 10^4$ cm$^{-1}$ is used to identify the absorber materials in thin film solar cells [26].

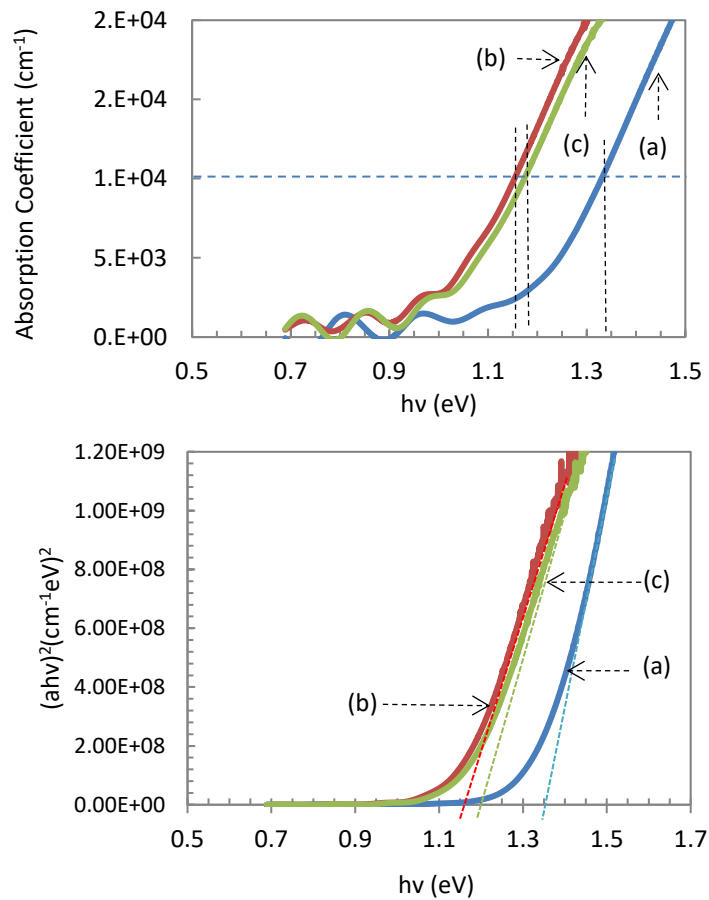

**Fig. 8:** Absorption coefficients $\alpha$ and $(\alpha h\nu)^2$ as a function of photon energy for the (a) as-deposited, (b) air-annealed and (c) $O_2$-annealed $Zn_3N_2$ films surfaces.

**Table 2:** Optical gaps and energy gap as absorption coefficient $\alpha = 10^{-4}$ of the as-deposited and annealed $Zn_3N_2$ films.

| Samples | As-deposited | Air-annealed | $O_2$-annealed |
|---|---|---|---|
| Optical gap | 1.35 | 1.15 | 1.20 |
| Energy gap | 1.38 | 1.15 | 1.18 |

*3.4 Electrical properties*

The electrical properties shown in Table 3 were measured by Van der Pauw Hall effect using Au as the contact electrodes under identical conditions. The film resistivity was very low and it was similar to the values reported by Futsuhara et al. [5]. As the films were annealed, more and more N atoms were activated. Therefore, more holes were produced, which resulted in the increase of the hole-concentration. When the film deposited for 30 min [22] was annealed in the air and $O_2$, N atoms were not enough to form N acceptors to compensate the electrons produced by zinc interstitials or oxygen vacancies, which mechanism transferred the p-type as-deposited film into n-type. However, for the films deposited for 1 hr., N atoms were enough to compensate the electrons and kept the conduction of the films as p-type. The mobility decreased with the hole-concentration increasing, which was like the phenomenon found by *Ji et al.* [29]. It can be interpreted as the scattering of holes generated by the ionized N acceptors in the annealed films.

**Table 3:** Resistivity, conduction type, mobility and carrier concentration (N) of the as-deposited and annealed $Zn_3N_2$ films. (Error of these measurements: Resistivity $\pm 0.001$ $\Omega$ and Mobility $\pm 0.1 cm^2/(V\cdot s)$).

| Samples | As-deposited | Air-annealed | $O_2$-annealed |
|---|---|---|---|
| Resistivity ($\Omega$.cm) | 0.77 | 2.0 | 1.9 |
| Conduction Type | p-type | p-type | p-type |
| Mobility [$cm^2/(V\cdot s)$] | 294.6 | 24.0 | 33.6 |
| N ($cm^{-3}$) x$10^{20}$ | 2.74 | 13.02 | 9.47 |

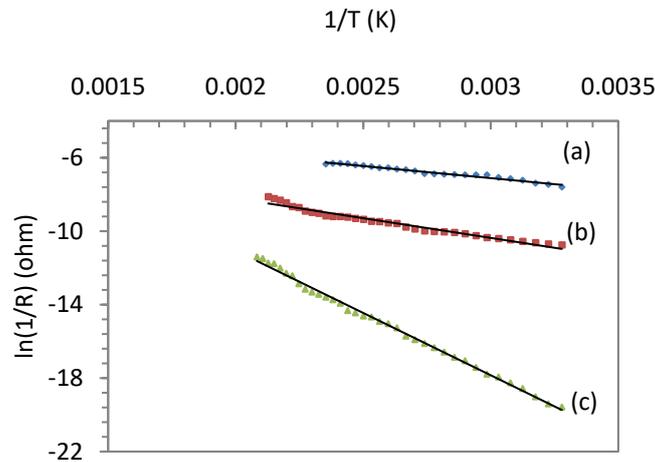

**Fig. 9:** Arrhenius plot of (a) as-deposited, (b) air-annealed, and (c) $O_2$-annealed $Zn_3N_2$ films.

Figure 9 showed the variation of logarithm of conductance against reciprocal of temperature for as-deposited and annealed $Zn_3N_2$ films. The measurement was carried out in vacuum with temperature varying from 20 °C to 250 °C. It was observed that the resistance of all films decreased as temperature increasing, which suggested semiconducting nature of $Zn_3N_2$ thin films. Furthermore, it was seen that the resistance of the annealed films was higher than that of the as-deposited film. The increased resistance in annealed films can be explained as the result of reduced carrier mobility shown in Table 3. The thermal activation energy of conductivity for the as-deposited, annealed $Zn_3N_2$ thin films was shown in Table 4. It was found that the activation energy was increased after annealing in the air and $O_2$. The likely reason was that the carriers with low mobility in the annealed films need more energy to be activated. Electronic excitation across this energy enables conduction between forbidden-gap energy levels, resulting in an increase in conductivity.

**Table 4**: Activation energy of the as-deposited and annealed $Zn_3N_2$ films.

| Samples | As-deposited | Air-annealed | $O_2$-annealed |
|---|---|---|---|
| Activation Energy (eV) | 0.12 | 0.19 | 0.58 |

*3.5 $Zn_3N_2$ based Photoconductor*

The photoconductor shown in Fig. 10 consists of a slab of $Zn_3N_2$ thin film with ohmic contacts affixed to the opposite ends. The films were irradiated with 850 nm light beam using an array of light emitting array (LED) with an intensity level of the surface 2.16 µW/mm².

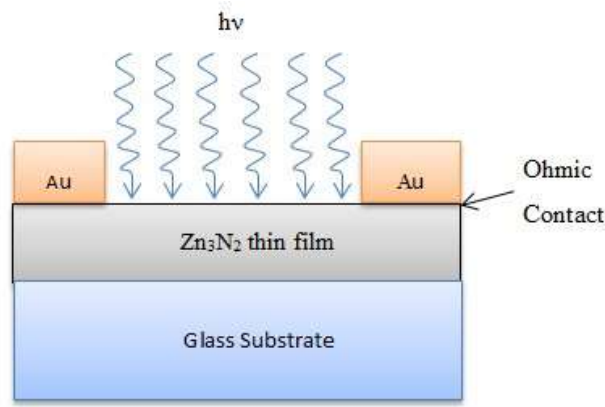

**Fig. 10:** Schematic diagram of the $Zn_3N_2$ photoconductor.

The long wavelength cut off is given by [30],

$$\lambda_c = \frac{hc}{E_g} = \frac{1.24}{E_g(eV)} \ (\mu m) \tag{5}$$

where, $\lambda_c$ is the wavelength corresponding to the $Zn_3N_2$ film band gap $E_g$ in Table 2. For wavelength shorter than $\lambda_c$, the incident radiation is absorbed by the film, and hole-electron pairs are generated by band-to-band transitions, resulting in an increase in conductivity. The intrinsic (band-to-band) photo excitation is illustrated in Fig. 11.

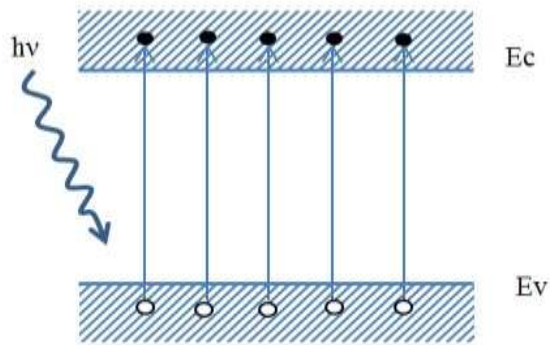

**Fig. 11:** Process of intrinsic photo excitation.

On the other hand, air and $O_2$ annealed films indicated very clear change of photoconductivity as shown in Fig.12 (a) and (b). Previous investigations also suggested that the oxygen played a critical role in electrical properties of zinc nitride films [24]. The decrease of the resistance was 1.39% and 1.02%, respectively. Comparing to the as-deposited film, the annealed films exhibited lower optical band gap and higher carrier concentration. It could be interpreted as both the intrinsic and extrinsic photo excitation easier occurred under the incident light.

The carrier mobility and optical band gap were close between the two annealed films. Nevertheless, the air-annealed film showed higher carrier concentration and lower activation energy. It indicated that besides the intrinsic excitation, the extrinsic photo excitation was more easily occurred than the $O_2$-annealed film. Therefore, the air-annealed film exhibited higher sensitivity than the $O_2$-annealed film. It was also noted that the photoconductivity was much higher than the corresponding films deposited for 30 min [22], which indicated that the photoconductivity was increased as increasing the deposition time.

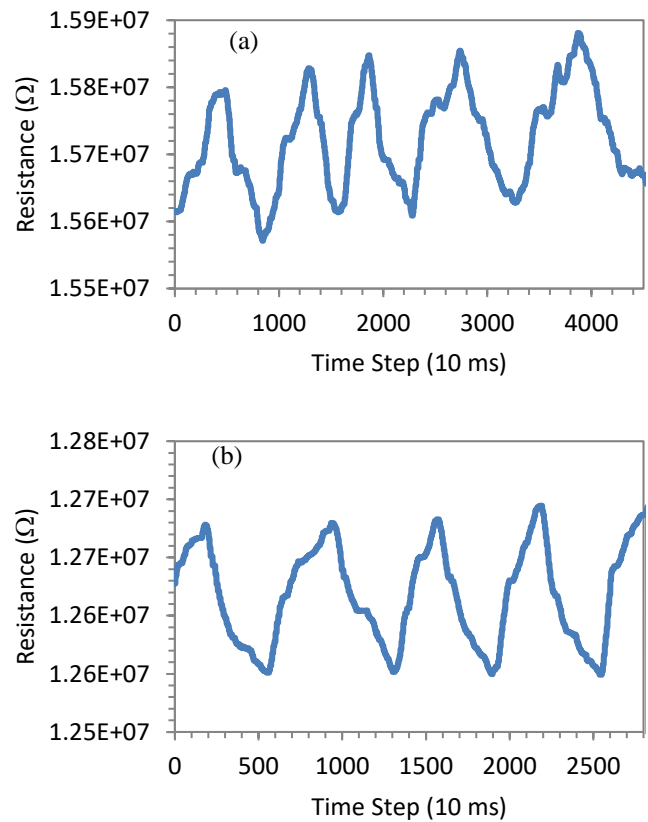

**Fig. 12:** Photo-response and photo-recovery for the $Zn_3N_2$ films annealed in the (a) air and (b) $O_2$. (Resolution of resistance measurement=0.005E+07 Ω).

The simulated solar light was+ converted into single wavelength between 300 nm to 1000 nm by *monochromator*. Fig. 13 illustrated the change of the resistance when the monochromatic lights fell on the surface of the air-annealed $Zn_3N_2$ film. It can be concluded that the film was photoconductive to all the monochromatic lights. The sensitivity was increased with the decreasing of the wavelength. The likely reason was that the light with lower wavelength generated more photon energy according to equation (5). Therefore, the more hole-electron pairs generated, the more resistance deceased.

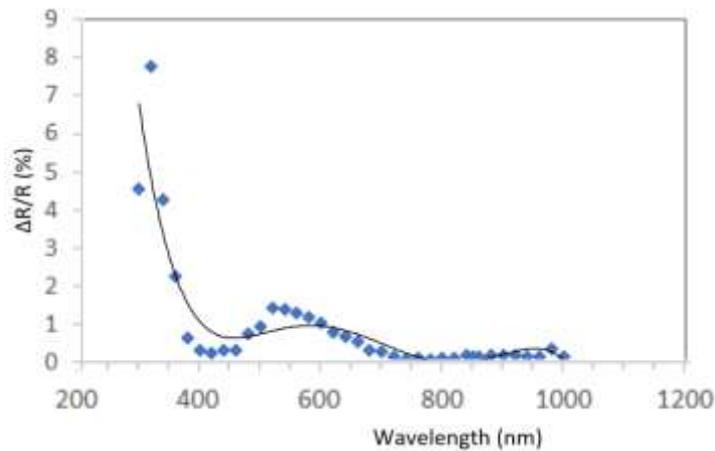

**Fig. 13:** Change of resistance as a function of wavelength for the $Zn_3N_2$ films annealed in the air. The curve was normalized to the spectral intensity of the lamp.

**4. Conclusion**

In this investigation, the $Zn_3N_2$ thin films were synthesized using the rf-magnetron sputtering and annealed in the air and $O_2$ gas mixture for 1 hr at 300 °C. The films were opaque, (400) phase preferred orientated and conductive p-type materials. The intensity of the predominant peak was dramatically improved after annealing in the air and $O_2$. Comparatively, the $Zn_3N_2$ films synthesized for 30 min were oriented at (321) phase and converted to n-type conduction after annealing. The calculated crystal size and surface roughness were not significantly changed by annealing. However, the calculated thickness of the film increased by ~35% with annealing. It can be interpreted as due to the improved surface crystalline structure and the interstitial oxygen ions introduced by the air and $O_2$. The annealing process reduced the optical band gap from 1.35 to 1.15 eV, which was close to that of the corresponding $Zn_3N_2$ films synthesized for 30 min. According to Van der Pauw measurement, the films exhibited low resistivity. Annealing process activated N acceptor, which decreased the mobility and increased the carrier concentration. The photoconductivity measurement showed that the as-deposited film was not irritated by the incident light. Nevertheless, the conductivity of the films annealed in the air and $O_2$ was increased by 1.39% and 1.02%, which is much higher than the corresponding films deposited for 30 min. Therefore, it can be concluded that longer deposition time heavily affected the structural, electric and optical properties of the $Zn_3N_2$ films, and improved the film photoconductivity in the end.

**Acknowledgements:** This work was supported by a grant (ECS-0928440) from National Science Foundation (NSF) of USA.